\documentclass[sigconf,authorversion,nonacm]{acmart}

\usepackage[utf8]{inputenc}

\usepackage{booktabs}
\usepackage{graphicx}
\usepackage[nolist]{acronym}
\usepackage{listings}
\usepackage{algorithm}
\usepackage{algorithmicx}
\usepackage{algpseudocode}
\usepackage[normalem]{ulem}
\usepackage[binary-units,abbreviations]{siunitx}
\DeclareSIUnit\Mbps{\mega\bit\per\second}
\usepackage{amsmath}
\usepackage{pgfplots}
\usepackage{tikzscale}

\usepackage{enumitem}
\usepackage[USenglish]{babel}

\AtBeginDocument{%
  \providecommand\BibTeX{{%
    \normalfont B\kern-0.5em{\scshape i\kern-0.25em b}\kern-0.8em\TeX}}}

\usepackage{ifthen}
\newboolean{cameraready}
\setboolean{cameraready}{true}

\newboolean{usetodonotes}
\setboolean{usetodonotes}{false}

\ifthenelse{\boolean{cameraready}}{
	
}{

}

\usepackage[textsize=scriptsize]{todonotes}
\ifthenelse{\boolean{usetodonotes}}{
	\setlength{\marginparwidth}{1cm}
	\paperwidth=\dimexpr \paperwidth + 6cm\relax
	\oddsidemargin=\dimexpr\oddsidemargin + 3cm\relax
	\evensidemargin=\dimexpr\evensidemargin + 3cm\relax
	\marginparwidth=\dimexpr \marginparwidth + 3cm\relax
	\presetkeys
	{todonotes}
	{noinline,caption={}}{}
}{
	\presetkeys
	{todonotes}
	{disable}{}
}

\begin{document}

\title[One Glitch to Rule Them All: Fault Injection Attacks Against AMD's SEV]{One Glitch to Rule Them All: Fault Injection Attacks Against AMD's Secure Encrypted Virtualization}

\author{Robert Buhren}
\email{robert.buhren@sect.tu-berlin.de}
\affiliation{
  \institution{Technische Universität Berlin - SECT}
  \country{}
}

\author{Hans Niklas Jacob}
\email{hnj@sect.tu-berlin.de}
\affiliation{
  \institution{Technische Universität Berlin - SECT}
  \country{}
}

\author{Thilo Krachenfels}
\email{tkrachenfels@sect.tu-berlin.de}
\affiliation{
  \institution{Technische Universität Berlin - SECT}
  \country{}
}

\author{Jean-Pierre Seifert}
\email{jpseifert@sect.tu-berlin.de}
\affiliation{
  \institution{Technische Universität Berlin - SECT}
  \institution{Fraunhofer SIT}
  \country{}
}

\renewcommand{\shortauthors}{Buhren, Jacob, Krachenfels, and Seifert}

\begin{abstract}

AMD \acf{sev} offers protection mechanisms for virtual machines in untrusted environments through memory and register encryption.
To separate security-sensitive operations from software executing on the main x86 cores, \ac{sev} leverages the \ac{amd-sp}.
This paper introduces a new approach to attack \ac{sev}-protected \acp{vm} by targeting the \ac{amd-sp}.
We present a voltage glitching attack that allows an attacker to execute custom payloads on the \acp{amd-sp} of all microarchitectures that support \ac{sev} currently on the market (Zen 1, Zen 2, and Zen 3).
The presented methods allow us to deploy a custom \ac{sev} firmware on the \ac{amd-sp}, which enables an adversary to decrypt a \ac{vm}'s memory.
Furthermore, using our approach, we can extract endorsement keys of \ac{sev}-enabled CPUs, which allows us to fake attestation reports or to pose as a valid target for VM migration without requiring physical access to the target host.
Moreover, we reverse-engineered the \acf{vcek} mechanism introduced with \ac{sev-snp}. %
The \ac{vcek} binds the endorsement keys to the firmware version of TCB components relevant for \ac{sev}.
Building on the ability to extract the endorsement keys, we show how to derive valid \acp{vcek} for arbitrary firmware versions.
With our findings, we prove that \ac{sev} cannot adequately protect confidential data in cloud environments from insider attackers, such as rogue administrators, on currently available CPUs.
\end{abstract}


\maketitle
\begin{acronym}
    \acro{ck}[cK]{component key}
    \acro{rk}[rK]{root key}
    \acro{ikek}[iKEK]{Intermediate Key Encryption Key}
    \acro{cbc}[CBC]{cipher block chaining}
    \acro{ecb}[ECB]{electronic codebook}
    \acro{iv}[IV]{initialization vector}
    \acro{psp}[PSP]{platform security processor}
    \acro{amd-sp}[AMD-SP]{AMD Secure Processor}
    \acro{smu}[SMU]{System Management Unit}
    \acro{vr}[VR]{voltage regulator}
    \acro{vid}[VID]{voltage identification}
    \acro{svi2}[SVI2]{serial voltage identification interface 2.0}
    \acro{ack}[ACK]{acknowledgement}
    \acro{tfn}[TFN]{telemetry function}
    \acro{sev}[SEV]{Secure Encrypted Virtualization}
    \acro{intelme}[IntelME]{Intel Management Engine}
    \acro{psp}[PSP]{Platform Security Processor}
    \acro{soc}[SoC]{system on a chip}
    \acroplural{soc}[SoCs]{systems on a chip}
    \acro{ftpm}[fTPM]{firmware TPM}
    \acro{vm}[VM]{virtual machine}
    \acro{sev-es}[SEV-ES]{SEV Encrypted State}
    \acro{sev-snp}[SEV-SNP]{SEV Secure Nested Paging}
    \acro{cs}[CS]{chip select}
    \acro{cek}[CEK]{chip-endorsement-key}
    \acro{ccp}[CCP]{crypto co-processor}
    \acro{pcb}[PCB]{printed circuit board}
    \acro{ic}[IC]{integrated circuit}
    \acro{pc}[PC]{program counter}
    \acro{csp}[CSP]{cloud service provider}
    \acro{sgx}[SGX]{Software Guard Extensions}
    \acro{asid}[ASID]{Address Space Identifier}
    \acro{pdh}[PDH]{Platform Diffie-Hellman Key}
    \acro{cek}[CEK]{Chip Endorsement Key}
    \acro{vcek}[VCEK]{Versioned Chip Endorsement Key}
    \acro{tcb}[TCB]{Trusted Computing Base}
    \acro{rmp}[RMP]{Reverse Mapping}
    \acro{em}[EM]{electromagnetic}
    \acro{spi}[SPI]{Serial Peripheral Interface}
    \acro{ma}[MA]{Migration Agent}
    \acro{oek}[OEK]{Offline Encryption Key}
    \acro{rom}[ROM]{read-only memory}
    \acro{svn}[SVN]{security version number}
    \acro{id}[ID]{256-bit identifier}
    \acro{ark}[ARK]{AMD Root Key}
    \acro{psp-os}[PSP OS]{PSP OS}
\end{acronym}

\acresetall
\section{Introduction\label{sec:intro}}
Introduced in 2016, AMD's \ac{sev} technology is the first commercially available solution aiming to protect \acp{vm} from higher-privileged entities~\cite{AMD2016}
Prominent use cases for \ac{sev} are cloud environments, where the high-privileged hypervisor has direct access to a \ac{vm}'s memory content.
In this scenario, a \ac{vm} without \ac{sev} is unprotected from an administrator with malicious intentions.
By encrypting a \ac{vm}'s memory, \ac{sev} aims to protect customers' data even when threatened by such an insider attack.

\begin{quote}
  \emph{``... SEV protects data-in-use enabling customer workloads to be protected cryptographically from each other as well as protected from the hosting software. Even an administrator with malicious intentions at a cloud data center would not be able to access the data in a hosted VM.''}~\cite[p. 9]{AMD2016}
\end{quote}

\noindent \Ac{sev} leverages AES encryption to ensure the confidentiality of data-in-use by transparently encrypting a \ac{vm}'s memory with a \ac{vm}-specific key.
The memory encryption is carried out by a dedicated memory encryption engine embedded in the memory controller~\cite{AMDSevAPI}. 
Recently presented extensions to \ac{sev}, \ac{sev-es} and \ac{sev-snp}, expand the encryption to the \ac{vm}'s register content and introduce software-based integrity protection using memory ownership tracking~\cite{sev-es,sev-snp}.
Besides runtime protection, \ac{sev} provides a remote attestation feature allowing \ac{vm}-owners to validate the correct instantiation of \acp{vm} even if the hypervisor is not fully trusted.

To ensure the confidentiality of \ac{vm} memory encryption keys and the integrity of the remote attestation feature, AMD CPUs contain a dedicated security co-processor, the \ac{amd-sp}\footnote{Formerly known as \ac{psp}}.
The \ac{amd-sp} constitutes the root-of-trust for modern AMD CPUs~\cite{ASPIntro} and manages \ac{sev}-related \ac{vm} life-cycle tasks such as deployment and migration~\cite{AMDSevAPI}.
The \ac{amd-sp} uses its own local memory and executes a firmware provided by AMD.
While the hypervisor, executing on the main x86 cores, is still in control of the \acp{vm}, i.e., it is responsible for scheduling \acp{vm}, only the \ac{amd-sp} can access a \ac{vm}'s memory encryption key.
This separation ensures that a malicious or compromised hypervisor cannot access a \ac{vm}'s data.

Previous research revealed that the \ac{amd-sp} is a single point of failure for the \ac{sev} technology~\cite{ccs19,CVE-2019-9836,buhren_uncover_2019}.
However, the presented issues are either limited to a specific CPU type, e.g., the issues presented in~\cite{ccs19,buhren_uncover_2019} only affect the first generation of AMD Epyc CPUs (Zen 1), or are effectively mitigated by firmware updates~\cite{CVE-2019-9836}.
To the best of our knowledge, no \ac{amd-sp}-related security issues that affect \ac{sev} are known for the current and last generation of AMD Epyc CPUs (Zen 2 and Zen 3).

Given the criticality of the \ac{amd-sp} for the security properties of the \ac{sev} technology, the question can be raised if there is a systematic way to mount attacks against \ac{sev}-protected \acp{vm} by targeting the \ac{amd-sp}.
Particularly, one could consider fault attacks that do not depend on the presence of software issues but instead force genuine code to enter an unintended state.
Recently, researchers applied this attack technique to Intel CPUs and mounted attacks against SGX enclaves~\cite{voltpillager,murdock_plundervolt_2020}. 

Due to its crucial role in the \ac{sev} technology, targeting the \ac{amd-sp} instead of the protected \acp{vm} potentially allows an attacker to circumvent any protection guarantees of \ac{sev}, independent from the targeted \ac{vm}.
Consequently, in this work, we answer the following research question: What are the implications of fault injection attacks against the \ac{amd-sp} for the \ac{sev} technology?

\subsection{Contributions}

In this work, we analyze the susceptibility of the AMD \ac{sev} technology towards physical attacks targeting the \ac{amd-sp}.
By manipulating the input voltage to AMD \acp{soc}, we induce an error in the \ac{rom} bootloader of the \ac{amd-sp}, allowing us to gain full control over this root-of-trust.

Building on this capability, we show that we can extract \ac{sev}-related secrets, i.e., \acfp{cek}, that can be leveraged to mount software-only attacks that do not require physical access to the target host.
Similar attacks have been previously presented in~\cite{ccs19}, however, in contrast to their approach, our attack does not depend on firmware issues and re-enables the attacks presented in their work on all \ac{sev}-capable CPUs.
Additionally, we reverse-engineered the new key-versioning scheme introduced by the \acf{sev-snp} extension that binds the \ac{cek} to TCB component versions.
This new key, called \ac{vcek}, is cryptographically bound to firmware versions of the target system.
Our glitching attack enables us to extract seeds for the \ac{vcek} that allow an attacker to derive the valid \acp{vcek} for all possible combinations of firmware versions.

We present our approach to determine the CPU-specific glitching parameters, i.e., the length and the depth of the voltage drop.
After determining these parameters in an initial characterization phase, our attack can be mounted fully automatic and requires no knowledge of the internal structure of the \ac{rom} bootloader.

Both the attack and the characterization of the target CPU require only a cheap (\textasciitilde30 \$) µController~\cite{teensy40} and a flash programmer (\textasciitilde12 \$), making this attack feasible even for attackers with no access to special equipment.
We successfully mounted the attack on AMD Epyc CPUs of all microarchitectures that support the \ac{sev} technology, i.e., Zen 1, Zen 2, and Zen 3.
We publish our firmware to mount the glitching attack, the code of our payloads and our implementation of \ac{sev}'s key-derivation functions under an open-source license at~\cite{pspReverseGlitch}.
To prove the successful extraction of endorsement keys, the repository includes valid signatures over the title of this paper.
The signatures can be validated using public keys, retrieved from AMD keyservers at~\cite{AMD_CEK_KDS, AMD_VCEK_KDS}.

We responsibly disclosed our findings to AMD, including our experimental setup and code.
AMD acknowledged our findings but refrained from providing an official statement regarding our attack.

\subsection{Overview\label{subsec:overview}}
In the following sections, we present our approach to overcome SEV's protection goals using voltage fault injection.

The presented attack allows an attacker to execute custom code on the AMD-SP by tricking the AMD-SP's ROM bootloader into accepting an attacker-controlled public key.
The AMD-SP uses this public key to validate the authenticity of firmware components, such as the SEV firmware.
The ability to execute code on the AMD-SP allows an attacker to a) exfiltrate confidential key material which impacts the entire SEV ecosystem's security and b) deploy a custom SEV firmware.

In Section~\ref{sec:background}, we present the required background information including information on: the SVI2 protocol, which is necessary to manipulate the AMD-SP's input voltage, the AMD-SP's firmware, including its protection mechanisms, and the SEV technology.
In Section~\ref{sec:attack_scenario}, we introduce two possible attack scenarios against SEV based on the attackers ability to execute code on the AMD-SP. Furthermore, we present our analysis of the AMD-SP's secure boot mechanism.
Our voltage glitch setup and attack approach is described in Section~\ref{sec:exp_setup}.
In Section~\ref{sec:firmware_decryption}, we describe our approach to decrypt firmware components of AMD Epyc Zen 3 systems to allow an analysis of the new \ac{vcek} key-derivation scheme introduced with \ac{sev-snp}, which is presented in Section~\ref{sec:key_derivation}.
Finally, we discuss the implications of the presented attacks in Section~\ref{sec:discussion} and conclude in Section~\ref{sec:conclusion}.

\section{Related Work\label{sec:related_work}}

Voltage glitching attacks targeting security-sensitive operations on CPUs have been subject to extensive analysis in the past.
The majority of reported attacks have been carried out on small embedded systems and \acp{soc}, where typically crowbar circuits (see Section~\ref{subsec:fi} for details) are used to inject the fault, e.g., in ~\cite{timmers_controlling_2016, galauner_glitching_2018, boone_there_2020}.
A more thorough list of voltage glitching attacks can be found in ~\cite{gianlucap_gipi_2021}.

More recently, voltage glitching attacks against Intel desktop and server CPUs have been reported, which use available interfaces to \acp{vr} for glitching the supply voltage.
\todo{More recently, voltage glitching attacks which use available interfaces to \acp{vr} for glitching the supply voltage, against Intel desktop and server CPUs have been reported.}
Several authors demonstrated how code running in the Intel SGX enclaves can be faulted by injecting glitches through a software-based voltage scaling interface~\cite{kenjar_v0ltpwn_2020, qiu_voltjockey_2020, murdock_plundervolt_2020}.
Thereby, SGX's integrity properties are violated, and keys from cryptographic operations running inside the secure enclave can be extracted.

The work most related to our attack is presented by Chen et al. in~\cite{voltpillager}.
The authors demonstrate the first physical attack targeting SGX enclaves entitled \emph{VoltPillager}.
VoltPillager improves the timing precision of software-based fault attacks and leverages direct hardware access to the \ac{vr} for injecting glitches.
By connecting wires to the bus between the CPU and the \ac{vr}, the authors could inject commands causing voltage glitches with higher timing precision than the previously mentioned software-based fault injection methods.
Furthermore, the attack is also applicable on patched systems, where the software interfaces for controlling the voltage are not accessible to an adversarial process.
Although our attack uses the same mechanism to alter the input voltage to the \ac{soc}, namely the external \ac{vr}, several factors distinguish our approach from theirs.
We, therefore, compare our approach with VoltPillager in Section~\ref{sec:attack_scenario}.\\

Since its introduction in 2016, several attacks against \ac{sev} have been published~\cite{Hetzelt:2017, du2017secure, morbitzer2018, morbitzer_extracting_2019,werner_severest_2019, li_exploiting_2019, wilke_sevurity_2020, severity_2021, underSErVed_2021}.
These attacks either rely on the ability to write to encrypted guest memory, the ability to access the guest's general-purpose register, or the ability to alter the mapping between guest-physical and host-physical pages of a \ac{sev}-protected \ac{vm}.
\Ac{sev-es} effectively mitigates attacks that require access to a guest's register state, and \ac{sev-snp} mitigates attacks that alter a guest's memory layout or content.
A different direction is explored in~\cite{radev_exploiting_2020}.
The authors present issues inside the Linux kernel of \ac{sev}-enabled guests that allow the circumvention of \ac{sev}'s security properties.
By manipulating the result of the \texttt{cpuid} instruction, they show how an attacker could trick the guest into not enabling the SEV protection at all.
To counter this issue, \ac{sev-snp} introduces a ``Trusted CPUID'' feature that prevents a hypervisor from reporting an invalid CPU feature set.

In~\cite{ccs19}, the authors analyze \ac{sev}'s remote attestation mechanism.
They revealed security issues in the \ac{amd-sp}'s firmware that enable an attacker to deploy a custom \ac{sev} firmware and extract keys critical to the remote attestation.
Using a manipulated \ac{sev} firmware, an attacker can override the debug policy of \ac{sev}-enabled \ac{vm}'s and thereby decrypt its memory.
The extracted keys can be used to fake the presence of \ac{sev} during deployment or migration.
These attacks require the presence of firmware issues in the \ac{amd-sp}.
Although the work shows that the \ac{amd-sp} is crucial for the security properties of \ac{sev}, the presented firmware issue is only present on the first generation of \ac{sev} capable CPUs (Zen 1). 
To the best of our knowledge, no comparable firmware issue for later generations of AMD CPUs (Zen 2 and Zen 3) has been published.

\section{Background\label{sec:background}}
This section introduces the \acf{sev} technology, voltage fault injection as means to induce errors in security-sensitive operations, and the \ac{vr} communication protocol.

\subsection{Secure Encrypted Virtualization\label{subsec:sev}}
The \ac{sev} technology offers protection mechanisms for \acp{vm} in untrusted environments, such as cloud environments~\cite{AMD2016}.
In contrast to Intel's \ac{sgx}, which focus on protecting \emph{parts} of an application, \ac{sev} protects full \acp{vm}.
The \emph{runtime protection} of \acp{vm} is achieved by transparently encrypting a \ac{vm}'s memory.
The \emph{remote attestation} feature of \ac{sev} allows cloud customers to validate the correct deployment of the \ac{vm}.
Since its introduction in 2016, AMD has introduced two extensions to \ac{sev} that add additional protection features to \ac{sev}.
While \ac{sev-es} adds encryption for guest \ac{vm} registers~\cite{sev-es}, \ac{sev-snp} introduces, amongst others, \emph{software-based} integrity protection and an enhanced \ac{tcb} versioning feature for the \acf{cek}~\cite{sev-snp}.
The \ac{cek} cryptographically links the target platform to the AMD root of trust.

Both the runtime protection and the remote attestation feature require the hypervisor to use an interface provided by a dedicated firmware running on the \ac{amd-sp}.
The API for \ac{sev} and \ac{sev-es} is specified in~\cite{AMDSevAPI}, whereas \ac{sev-snp} uses a dedicated API specified in~\cite{AMDSNPAPI}. 
The \ac{sev} firmware is responsible for managing the memory encryption keys of the \acp{vm} and implementing the remote attestation feature of \ac{sev}. 

Our approach re-enables the previously presented attacks by Buhren et al. by allowing the execution of custom code on the \ac{amd-sp}.
Therefore, for details on \ac{sev} and \ac{sev-es} protected systems, we refer to their paper~\cite{ccs19}, while the remainder of this section focuses on the \ac{sev-snp} technology.

\subsubsection{SNP Runtime Protection\label{subsubsec:snp_runtime_protection}}
In addition to the memory encryption introduced by \ac{sev} and the register encryption introduced by \ac{sev-es}, \ac{sev-snp} adds software-based memory integrity protection.
For \ac{sev-snp} enabled \acp{vm}, the CPU will track ownership of memory pages using the \ac{rmp} table.
Memory accesses are subject to an \ac{rmp} check to ensure that, e.g., the hypervisor cannot access encrypted guest memory or manipulate the mapping between guest-physical and host-physical.
The \ac{rmp} access check mitigates previously presented attacks that rely on the hypervisor's ability to write or remap a \ac{vm}'s memory.

\subsubsection{SNP Remote Attestation\label{subsubsec:snp_remote_attestation}}
With \ac{sev-snp}, a \ac{vm} can request an attestation report at an arbitrary point in time.
To that end, the \ac{vm} communicates with the \ac{amd-sp} via an encrypted and integrity-protected channel.
The \ac{sev} firmware will generate an attestation report that includes a measurement of the initial \ac{vm} state and additional information about the host platform.
A \ac{vm} can also include 512 bits of arbitrary data in the report, e.g., a hash of a public key generated in the \ac{vm}.
The \ac{vm} can then provide this attestation report to a third party, such as the guest owner.
The attestation report is signed with platform specific endorsement key, the \ac{vcek}.
Using an ID provided by the \ac{amd-sp}, a guest owner can retrieve a signed \ac{vcek} for a specific AMD \ac{soc} from an AMD key server~\cite{AMD_VCEK_KDS}.
The \ac{vcek} is signed by the \acf{ark} which can be retrieved from an AMD website~\cite{AMD_MILAN_ARK}.
For each AMD Zen microarchitecture, there exists a different \ac{ark}.
Using the obtained \ac{vcek} and the \ac{ark}, the guest owner can validate that an authentic \ac{amd-sp} has issued the report.
The signed attestation report links the data in the report provided by the guest to the respective \ac{vm}.
If the \ac{vm} provided the hash of a public key within the attestation report, a genuine report proves that the \ac{vm} owns the corresponding key pair.

\subsubsection{SNP Versioned Chip Endorsement Key\label{subsubsec:snp_vcek}}
\Ac{sev} and \ac{sev-es} rely on a static, non-revocable ECDSA key (the \ac{cek}), to authenticate a remote AMD \ac{soc}.
Firmware issues that enable \ac{cek} extraction, as presented in~\cite{ccs19}, have severe implications for \ac{sev}.
An extracted \ac{cek} allows an attacker to fake attestation reports or pose a valid target for \ac{vm} migration.
As it is impossible to revoke a \ac{cek}, firmware updates are not sufficient to mitigate these attacks.

\Ac{sev-snp}, therefore, introduces the \acf{vcek}. 
A \ac{vcek} is derived on the \ac{amd-sp} from chip-unique fused secrets and bound to firmware security versions of components which are part of \ac{sev}'s TCB.
These \acfp{svn} are combined in a single TCB version string as shown in Table~\ref{tab:tcb_version}.
\begin{table}[h!]
\begin{tabular}{@{}lllllllll@{}}
\toprule
Byte(s)  & 0          & 1 & \multicolumn{4}{l}{2-5} & 6 & 7     \\ \midrule
Field & BOOT\_LOADER & TEE  & \multicolumn{4}{l}{RSVD}  & SNP   & MICROCODE \\ \bottomrule
\end{tabular}
\vspace{.5em}
\caption{\Ac{sev-snp}'s TCB version string~\cite[Chapter 2.2.]{AMDSNPAPI}.}
\label{tab:tcb_version}
\vspace{-1em}
\end{table}
In case of a known firmware issue, an update of a single TCB component will result in a different \ac{vcek}.
To retrieve the signed \ac{vcek}, the user has to provide the ID of the target platform, as well as the \acp{svn} for which the signed \ac{vcek} should be retrieved.

\Ac{sev-snp} allows to downgrade the \ac{svn} of the TCB components to provide backward compability.
To that end, \ac{sev-snp} provides the \texttt{SNP\_SET\_REPORTED\_TCB} API call.
The firmware ensures that the call can only be used to set a lower TCB version.
Providing higher \acp{svn} than the current counter values results in an error.

In contrast to the \ac{cek}, the \ac{vcek} is cryptographically bound to specific firmware versions.
Hence, previously extracted \acp{vcek} are no longer valid after a firmware upgrade.
Any party involved in the attestation process can now enforce  minimum TCB component versions.

\subsubsection{SNP Migration\label{subsubsec:snp_migration}}
\noindent Migration of \ac{sev}-protected \acp{vm} requires a dedicated mechanism, as the \ac{vm} memory encryption key is solely accessible by the \ac{amd-sp}.
For \ac{sev} and \ac{sev-es}, the \ac{amd-sp} is involved in the migration processes and policy enforcement.
Using a Diffie-Hellman key exchange, the involved \acp{amd-sp} on the source and target of the migration derive shared transport keys to migrate the memory.

To allow more complex migration schemes, \ac{sev-snp} introduces \acp{ma}.
A \ac{ma} is a dedicated \ac{vm} associated with one or multiple \acp{vm} and is responsible for migrating a \ac{vm}.
In the first step, the hypervisor uses the \texttt{SNP\_PAGE\_SWAP\_OUT} \ac{sev-snp} API command to export a \ac{vm}'s memory.
The \ac{amd-sp} will re-encrypt the memory using a dedicated key, the \ac{oek}.
The \ac{amd-sp} generates the \ac{oek} during the initial launch of a \ac{vm}.

Then the hypervisor calls the \ac{ma}, which will retrieve the \emph{guest context} of the respective \ac{vm} using the \texttt{VM Export} \ac{amd-sp} API command.
The context represents the internal \ac{vm} state for \ac{sev-snp} and contains, amongst others, the \ac{oek} used to re-encrypt a \ac{vm}'s memory pages during migration.
The \ac{amd-sp} ensures that the context is only exported to \acp{ma} that are associated with the respective \ac{vm}.
The \ac{ma} can now enforce arbitrary policies for the migration process, as only the \ac{ma} can decrypt the memory pages.
To re-import a \ac{vm}, a \ac{ma} on the target host can re-create the \ac{vm} using the guest context and the encrypted guest memory.

The \ac{ma} associated with a \ac{vm} is part of a \ac{vm}'s TCB, as it can retrieve the guest context including the \ac{oek}.
To enable guest owners to validate the \ac{ma} associated with their \ac{vm}, the \ac{amd-sp} remote attestation reports include the measurement of the \ac{ma}.

Alternatively, \ac{sev-snp} supports a guest-assisted migration mode where the memory pages are transferred by trusted component within the guest itself.

\subsection{AMD Secure Processor\label{subsec:amd-sp}}
Initially introduced in 2013 under the name \acf{psp}~\cite{lai_psp_intro}, the \ac{amd-sp} is a dedicated security processor and contained within AMD CPUs.
The \ac{amd-sp} is an ARMv7 core with dedicated SRAM executing a firmware provided by AMD and is the root-of-trust for the AMD \ac{soc}.
The \ac{amd-sp} executes a firmware that implements the \ac{sev}-related functions defined in the \ac{sev}-API~\cite{AMDSevAPI}, respectively the \ac{sev-snp}-API~\cite{AMDSNPAPI}.
The firmware is loaded from an SPI-attached flash chip and is stored alongside the UEFI firmware~\cite{buhren_uncover_2019}.

\paragraph{AMD-SP Boot Procedure}

\begin{figure}
  \includegraphics[width=0.75\columnwidth]{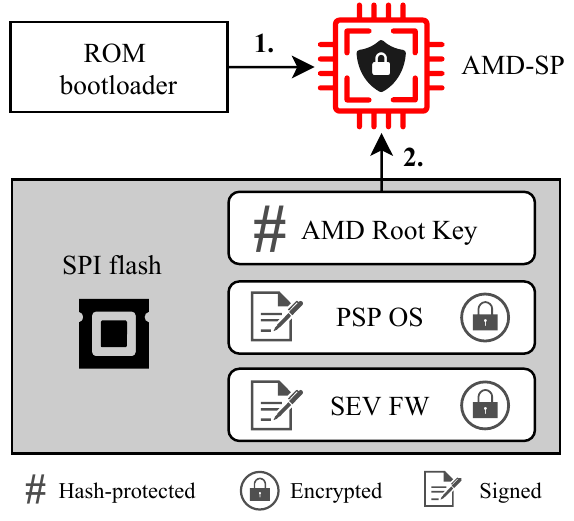}
  \vspace{-2mm}
  \caption{Overview of the \ac{amd-sp}'s firmware components relevant to the TCB of \ac{sev} protected \acp{vm}.}
  \label{fig:amd_sp_boot}

\end{figure}

\cite{ccs19} analyzes the \ac{amd-sp}'s boot procedure on AMD Epyc Zen 1 CPUs.
Figure~\ref{fig:amd_sp_boot} depicts \ac{amd-sp}'s firmware components relevant for \ac{sev}.
On these systems, the \ac{amd-sp} initially starts executing from a non-updatable \ac{rom} bootloader, see Figure~\ref{fig:amd_sp_boot} Step 1.
The \ac{rom} bootloader is responsible for loading and verifying an RSA public key from a modifiable SPI flash. 
This public key is used to validate the integrity of files loaded from the SPI flash.
The public key itself is verified using hashes stored within the bootloader \ac{rom}, Step 2 of Figure~\ref{fig:amd_sp_boot}.

In the following steps, the \ac{rom} bootloader loads another boot stage, called the \emph{PSP OS\acused{psp-os}} by Buhren et al.~\cite{ccs19}, from the SPI attached flash.
This boot stage contains a proprietary operating system and will later load and verify the \ac{sev} firmware from flash.
Both this second boot stage and the \ac{sev} firmware are validated using the public key loaded by the \ac{rom} bootloader.
The public key used to authenticate the PSP OS and the \ac{sev} firmware is identical to the \ac{ark} of the corresponding microarchitecture.

We confirmed that the described boot procedure is present in all CPUs we analyzed.
However, on AMD Epyc Zen 3 systems, both the PSP OS, as well as the SEV firmware component are encrypted and the \ac{sev} firmware is validated using a public key embedded in the PSP OS instead of the \ac{ark}.
In Section~\ref{sec:firmware_decryption}, we describe how we decrypt these components to enable further analysis.

\subsection{Fault Injection by Voltage Glitching\label{subsec:fi}}
\todo[disable]{Thilo}

\Acp{ic} need to be operated under the specified conditions to function as intended, e.g., within rated supply voltage, clock stability, temperature, and electromagnetic field ranges~\cite{bar-el_sorcerer_2006}.
This dependency can be misused to force faulty behavior during the chip's operation.
Glitches on the supply voltage line, i.e., short supply voltage variations, can be used to produce computational errors on CMOS circuits at low cost~\cite{djellid-ou_supply_2006}.
Unintended bit flips, corrupted instructions, and skipping of instructions in a microprocessor are examples of such errors.
If these errors are forced during the execution of cryptographic algorithms, information about the secret key or plaintext might be leaked~\cite{bar-el_sorcerer_2006}.
On the other hand, faults can be used to skip security checks, enter protected code paths, or gain code execution~\cite{timmers_controlling_2016, lu_injecting_2019}.

Depending on the design of the target, different approaches can be used to inject faults into the supply voltage rail.
In case the voltage is supplied externally to the \ac{pcb}, an external power supply can introduce glitches through that interface.
If the voltage is generated directly on the \ac{pcb} using a \acf{vr}, the injection of glitches becomes more complex.
On the one hand, glitches can be injected using a so-called crowbar circuit, which creates a short circuit between the voltage line and GND, effectively enforcing a voltage drop~\cite{oflynn_fault_2016}.
On the other hand, on more advanced systems, such as \acp{soc}, the \acp{vr} typically offer communication interfaces to adjust the voltage on demand.
These interfaces, if not adequately protected, can also be leveraged to inject voltage glitches~\cite{murdock_plundervolt_2020, qiu_voltjockey_2020, voltpillager}.

\todo[disable]{
Literature:
\begin{itemize}
	\item Shaping the Glitch: Optimizing Voltage Fault Injection Attacks \cite{bozzato_shaping_2019}
\end{itemize}
}

\subsection{SVI2 Protocol\label{subsec:svi2}}
\todo[inline, disable]{Thilo}

The demand for processors trimmed for high performance which at the same time show deterministic behavior, has put increased requirements on the power management of x86 processors~\cite{advancedm_white_2018}.
The power consumption of a processor is directly linked with its current consumption and supply voltage.
To maximize performance gain, the power consumption in modern processors is managed by dedicated on-chip µControllers, which measure voltage/current in real time. \todo{the vr isn't a µController!?}
Recent AMD processors dynamically monitor and adjust their primary (Core and SOC) voltage rails, which is also known as dynamic voltage scaling~\cite{advancedm_white_2018}.
Through the \acf{svi2}, the processor can directly communicate with a \ac{vr} to monitor and alter the supply voltages.
The AMD \ac{svi2} is a three-wire interface with clock (SVC), data (SVD), and telemetry (SVT) lines.
Although the corresponding  specification by AMD is not publicly available, all the necessary information on \ac{svi2} can be gathered from datasheets of different \acp{vr} implementing the interface, e.g., from \cite{internatio_ir35201_2015, internatio_ir35204_2016, richtekte_dualoutput_2019, renesasel_isl62776_2020}.

\todo{Niklas: You can also set both voltages (Core and/or \ac{soc})}
\todo{Thilo: Is that important?}
The \ac{svi2} protocol is similar to the I\textsuperscript{2}C bus concept.
The CPU acts as master and sends control packets via the SVC and SVD lines to the \ac{vr}.
\Ac{svi2} control packets consist of 3 bytes transmitted conforming to the \emph{SMBus send byte} protocol: 1 byte for selecting the voltage domain (Core or SOC) followed by an \ac{ack} bit, and then 2 bytes containing the voltage to be applied and other configuration parameters, each byte followed by an \ac{ack} bit~\cite{renesasel_isl62776_2020}.
Due to the configuration encoding, the voltage can be configured with a step size of 6.25\,mV.
Through the \ac{tfn} configuration bits, periodic voltage (and current) reports from the \ac{vr} to the CPU via the SVC and SVT lines can be enabled.
Details about the telemetry package format can be found in~\cite{richtekte_dualoutput_2019}.

\section{Attack Scenario\label{sec:attack_scenario}}

One of the most prominent use cases for the \ac{sev} technology, are cloud environments.
In cloud environments, the physical systems hosting the \acp{vm} are under full control of a \acf{csp}.
In our attack scenarios, the attacker aims to access a \ac{sev}-protected \ac{vm}'s memory content by attacking the \ac{amd-sp}.
We make no assumptions on whether \ac{sev-es}, \ac{sev-snp}, or just \ac{sev} is active.
We consider an attacker who has either access to the physical hosts that execute the targeted \ac{vm} or access the \ac{csp}'s maintenance interfaces that allow to, e.g., migrating a \ac{vm} to another physical system.
Examples for attackers with these capabilities are maintenance or security personnel or system administrators of the \ac{csp}.
We do not assume the presence of firmware or software bugs in the targeted host or \ac{vm} for our attack scenarios.
Based on these capabilities, we showcase two different approaches to access a \ac{sev}-protected \ac{vm}'s data.
The attack scenarios are inspired by the attacks presented in~\cite{ccs19}, but are adapted to \ac{sev-snp}.
We want to emphasize that these scenarios are merely two examples of possible attacks.
Due to the \ac{amd-sp}'s critical role for the \ac{sev} technology, targeting the \ac{amd-sp} potentially enables several other attack scenarios.

\paragraph{Scenario 1: Debug Override} As previously presented in~\cite{ccs19}, the \ac{sev} API provides debug features that allow the de- and encryption of a \ac{vm}'s memory~\cite[Chapter 7]{AMDSevAPI}.
A similar feature exists for \ac{sev-snp}~\cite[Section 8.23]{AMDSNPAPI}.
Both \ac{sev}'s and \ac{sev-snp}'s debug features are subject to a policy check enforced by the \ac{sev} firmware.
Only if a guest owner explicitly enabled debugging during the initial deployment, the \ac{sev} firmware will allow the debug API commands.

By altering the \ac{sev} firmware, an attacker could override this policy enforcement to allow the debug commands regardless of a guest owner's policy.
To that end, the attacker must replace the \ac{sev} firmware on the physical machine that hosts the target \ac{vm}.
Alternatively, the attacker could first migrate the targeted \ac{vm} to a previously prepared system.
The attacker can then use the previously mentioned debug API calls to decrypt a \ac{vm}'s memory regardless of the policy specified by the guest owner.

\paragraph{Scenario 2: Forge Attestation} In this second scenario, the attacker has access to the control interface of the hypervisor to initiate the migration of \ac{sev}-protected \acp{vm}.
However, in contrast to the first scenario, the attacker does not need to alter the firmware of the targeted host; hence no physical access to the targeted host is required.
Instead, the attacker needs to extract CPU-specific endorsement keys of an \ac{sev}-capable CPU to sign arbitrary \ac{sev} attestation reports.
These endorsement keys play a central role in the remote attestation feature of the \ac{sev} technology, see Section~\ref{subsubsec:snp_runtime_protection}.
To decrypt a \ac{vm}'s memory of an \ac{sev-snp}-protected \ac{vm}, the attacker fakes the attestation report during deployment or migration to trick a \ac{vm} owner into accepting a malicious \ac{ma}.
The \ac{ma} is part of a \ac{vm}'s TCB and has access to the \acf{oek} of \acp{vm}, see Section~\ref{subsubsec:snp_migration}.
Using the \ac{oek}, a malicious \ac{ma} can decrypt a \ac{vm}'s memory.

For pre-\ac{sev-snp} systems, the \ac{sev} firmware is responsible for handling migration.
As the pre-\ac{sev-snp} firmware will only accept endorsement keys of the same microarchitecture, the attacker has to extract an endorsement key of a CPU from the same microarchitecture as the targeted host's CPU.
In other words, to attack a \ac{vm} running on a Zen 2 CPU, the extracted endorsement keys must also belong to a Zen 2 CPU.
Furthermore, pre-\ac{sev-snp} systems might require the endorsement keys to be signed by the host owner's certificate authority, e.g., the CA of the \ac{csp}.
In this case, the attacker must be able to acquire a valid signature from the CA for the extracted endorsement keys.
This procedure is also required when integrating a new \ac{sev}-capable system in an existing cloud infrastructure and can be seen as part of a \ac{csp} administrator's responsibilities.
For Zen 1 systems, the migration attack was previously presented in~\cite{ccs19}.\\

Both presented example scenarios require the attacker to gain code execution on the \ac{amd-sp}.
Therefore, in the following sections, we present our analysis of the \ac{amd-sp}'s susceptibility towards voltage fault injection as means to execute attacker-controlled code.

\subsection{Targeting the \ac{amd-sp}} \label{sec:attack_scenario:targeting_amd_sp}
\begin{figure}
  \includegraphics[width=\linewidth]{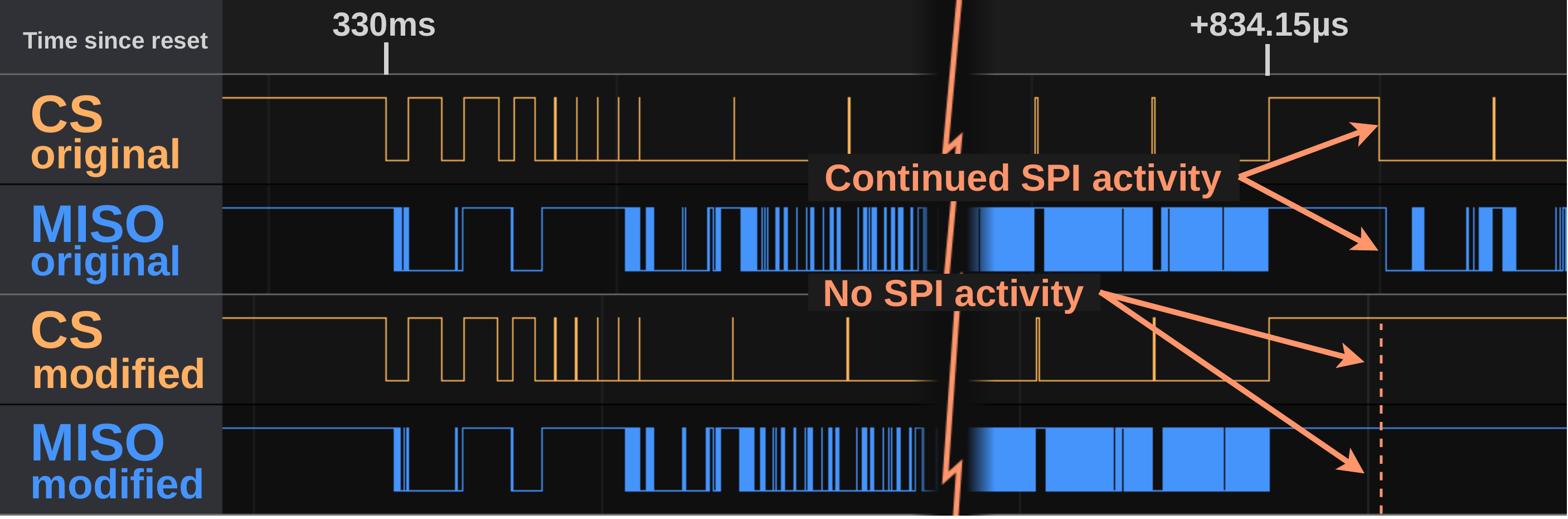}
  \caption{SPI bus traces during the initial boot. \acs{cs} and MISO lines only. The upper part depicts \acs{spi} bus activity for the original flash image (``\acs{cs} original'' and ``MISO original''). The lower part shows the corresponding \acs{spi} signals for a flash image with a manipulated \ac{ark}.}
  \label{fig:ark_spi}

\end{figure}

For the attack scenarios presented in the previous section, the attacker needs to execute custom code on the \ac{amd-sp}, either to provide a custom \ac{sev} firmware, or to extract the endorsement keys.
As described in Section~\ref{sec:background}, the \ac{amd-sp} loads an RSA public key, the \ac{ark}, from the SPI attached flash to validate the authenticity of subsequent loaded firmware components.
If an attacker would be able to replace the original \ac{ark}, all firmware components would be validated using the attacker-controlled key, thereby enabling the attacker to execute code directly after the \ac{rom} bootloader stage.

To better understand the \ac{ark} verification, we analyzed the traffic on the \ac{spi} bus during the boot process of an AMD Epyc CPU.
We conducted two experiments: first we recorded the \ac{spi} traffic during a normal boot, i.e., a boot with the original flash content.
The upper part of Figure~\ref{fig:ark_spi} shows the activity on the \ac{cs} and MISO lines of the SPI bus for this first experiment.

In a second experiment, we flipped a single, non-functional bit of the \ac{ark}. 
While the flipped bit would still allow validating signatures, the hash comparison by the \ac{rom} bootloader would fail.
The corresponding trace is shown in the lower part of Figure~\ref{fig:ark_spi}.
The \ac{cs} signal will be pulled low if the \ac{spi} master, in our case the \ac{amd-sp}, transmits data on the bus; otherwise the \ac{cs} signal is high.

Our analysis revealed a small period of time after the \ac{ark} is loaded without \ac{spi} traffic.
As we could not observe further \ac{spi} traffic when providing a manipulated \ac{ark}, we inferred that the \ac{amd-sp} validates the \ac{ark}'s integrity during this window.
Furthermore, we could observe that the amount \ac{cs} line changes prior to this gap only depends on the \ac{ark} size.

We identified this time period as a promising window of opportunity to inject our fault due to the following reasons:
\begin{itemize}
  \item Injecting a fault during the validation of the \ac{ark} potentially enables us to coerce the \ac{amd-sp} into accepting our own public key. By re-signing the flash image, we can manipulate all existing firmware components signed with that key.
\item The \ac{ark} validation happens at an early stage in the \ac{amd-sp}'s boot process. The fault injection might render the target system non-responsive which forces us to reset the target. By focusing on a very early security check we increase the number of glitches we can inject.
\item According to our observation, the amount of \ac{spi} traffic prior to the \ac{ark} validation only depends on the size of the \ac{ark}. This enables us to leverage the SPI traffic as a trigger for our fault injection.

\end{itemize}

\noindent To inject a fault during the \ac{ark} validation, we chose a similar approach as presented by Chen et al. in their attack called \emph{VoltPillager}~\cite{voltpillager}.
We explain the similarities and differences to their approach in the following section.

\subsection{Glitching the \ac{amd-sp}} \label{sec:attack_scenario:glitching_amd_sp}
To inject a fault, we leverage a CPU-external \ac{vr} to manipulate the input voltage of AMD \acp{soc}.
The \ac{vr} is an external controller that communicates via a dedicated bus, the \ac{svi2} bus, with the AMD \ac{soc} to allow the \ac{soc} to dynamically change the input voltage, e.g., when CPU-frequency changes require a different input voltage.
Our analysis of the AMD \ac{svi2} bus revealed that the external \ac{vr} not only controls the input voltage of the main x86 cores, but also the input voltage of the \ac{amd-sp}.
Although the \ac{svi2} protocol allows a single \ac{vr} to handle both input voltages, we observed that AMD Epyc systems leverage two independent \acp{vr} to handle the input voltages.
As described in Section~\ref{subsec:svi2}, the AMD \ac{soc} uses two different voltage domains, Core and \ac{soc}.
We verified that we can manipulate the \ac{amd-sp}'s input voltage via the \ac{soc} voltage domain.
Using a similar hardware setup as presented by Chen at al.~\cite{voltpillager}, we injected our own packets into the \ac{svi2} bus leveraging a \emph{Teensy} µController.

However, in contrast to the approach taken by Chen et al., where the authors target the protected entity, i.e., code executing in the SGX enclave, we target the \ac{amd-sp}.
To overcome the protections imposed by \ac{sev}, targeting the \ac{amd-sp} instead of the \ac{sev}-protected \ac{vm} has several benefits for the attacker:
\begin{itemize}
  \item System stability - If our fault attack renders the target unusable, we can simply reset the target and try again as the \ac{amd-sp} rom-bootloader will immediately execute once the \ac{soc} powers on. We don't need to fully instantiate the \ac{sev}-protected \ac{vm}.
  \item Attack effectiveness - Once our fault injection is successful, the decryption of \ac{vm} memory is 100\% effective.
  \item Independence from target \ac{vm} - Our approach works for all \ac{sev}-protected \acp{vm}, regardless of the type of operating system or application used inside the \ac{vm}.
  \item Key extraction - Targeting the \ac{amd-sp} allows us to extract \ac{sev}-related secrets which can be used to target remote systems. For these systems, we don't require physical access.
  \item Automation of the attack - Once the target CPU is characterized, i.e., the glitching parameters are determined, subsequent attacks require no manual intervention.
  \item Blinded glitching - Our glitching attack solely relies on observing an \emph{external} trigger, the \acf{cs} signal of the \ac{spi} bus. We don't require code execution on the target to determine our glitching parameters.
\end{itemize}

\noindent In the following section we present our experimental setup that allows us to inject faults into the \ac{amd-sp}'s \ac{rom} bootloader.

\section{Glitch Attack} \label{sec:exp_setup}
\begin{figure}[tb]
    \includegraphics[width=\columnwidth, trim=0 0 0 0, clip]{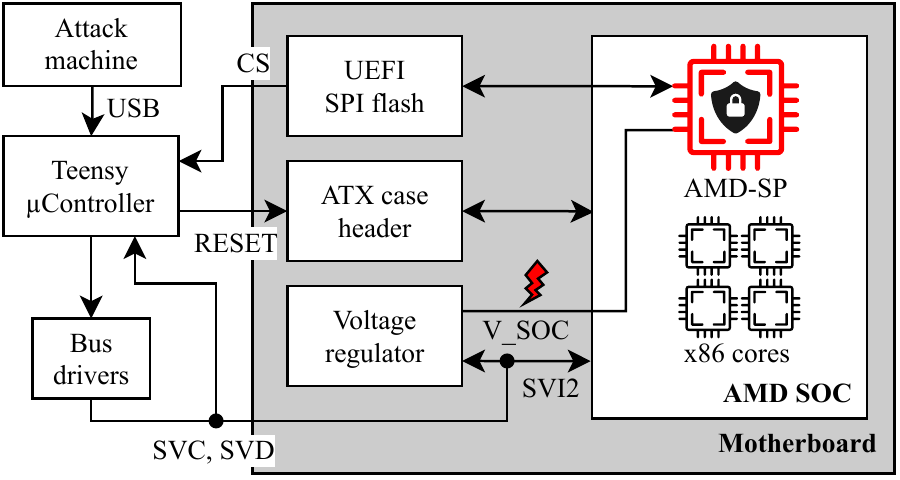}
    \vspace{-5mm}
    \caption{Schematic of the attack setup.}
    \label{fig:hardware-setup}
\end{figure}

To overcome the boot protection mechanisms of the \ac{amd-sp}, we target the \ac{rom} bootloader's signature verification of the \ac{ark} with our glitching attack.
Figure~\ref{fig:hardware-setup} depicts our glitching setup and the components involved.
Inspired by the \emph{Voltpillager} attack \cite{voltpillager}, we use a Teensy 4.0 \si{\micro}Controller \cite{teensy40} for all communication with the low-level hardware and to run the time-critical attack logic.
The Teensy is responsible for monitoring the \acf{cs} line of the target motherboard's SPI bus to identify the precise time to perform the glitch and whether a glitch was successful or not, see Section~\ref{sec:attack_scenario:targeting_amd_sp}.
In order to drop the voltage of the \ac{amd-sp}, the Teensy is connected to the \ac{svi2} bus of the target.
By injecting packets into this bus, the Teensy programs the \ac{vr} to apply the corresponding voltage levels.
For resetting the target \ac{soc} after a failed attack, the Teensy is connected to the ATX Reset line.

The Teensy is controlled from an attack machine via a serial-over-USB interface.
This attack machine is responsible for selecting attack parameters and orchestrating the glitching attacks.
We want to emphasize that the Teensy µController is capable of performing the attack on its own with only minor firmware modifications.

Using the described setup, we were able to successfully execute custom payloads on the CPUs shown in Table~\ref{tbl:cpus}.
We used the \emph{Supermicro H11DSU-iN} motherboard\footnote{Although the H11DSU-iN does not officially support the 72F3 CPU, we still could successfully boot the \ac{amd-sp}.} for all targeted CPUs.
In the following sections, we describe the required steps to mount our glitching attack.
\begin{table}[h]
\begin{tabular}{cccc}
CPU & \si{\micro}Architecture   & Previously Exploited  \\
\hline
72F3    & Zen 3 (Milan)             & No \\
7272     & Zen 2 (Rome)               & No      \\
7281     & Zen 1 (Naples)              & Yes \cite{ccs19}
\end{tabular}
\vspace{1mm}
\caption{\acp{amd-sp} successfully attacked.\label{tbl:cpus}}
\vspace{-2em}
\end{table}

\subsection{Payload Preparation\label{sec:exp_setup:payload_preparation}}
As a pre-requisite for our attack, we prepare the \ac{spi} flash image of the target so that our payload replaces the \ac{psp} OS component in the target's flash image, see Section~\ref{subsec:amd-sp}.
Then we replace the \ac{ark} with our own public key and re-sign the payload with this key.
In case of a successful glitch, the \ac{amd-sp} accepts our public key and executes our payload instead of the original PSP OS component.
As a proof-of-concept payload, we use a simple ``Hello World'' application, which outputs the string ``Hello World'' on the \ac{spi} bus.
After the attack, we can verify that we gained code execution by reading ``Hello World'' from the \ac{spi} bus using a logic analyzer.

\subsection{Attack Cycle} \label{sec:exp_setup:attack_cycle}

To coerce the \ac{amd-sp} into accepting our public key, we need to inject a fault during the hash verification of the \ac{ark}.
The attack can be split into several steps, executed in a loop until a successful glitch was detected.
For each targeted CPU, we first determine static glitch parameters: \emph{delay} and \emph{duration}.
In Section~\ref{sec:exp_setup:parameters}, we explain our approach for identifying these parameters in detail.

\begin{figure*}[t]
    \includegraphics[width=\textwidth]{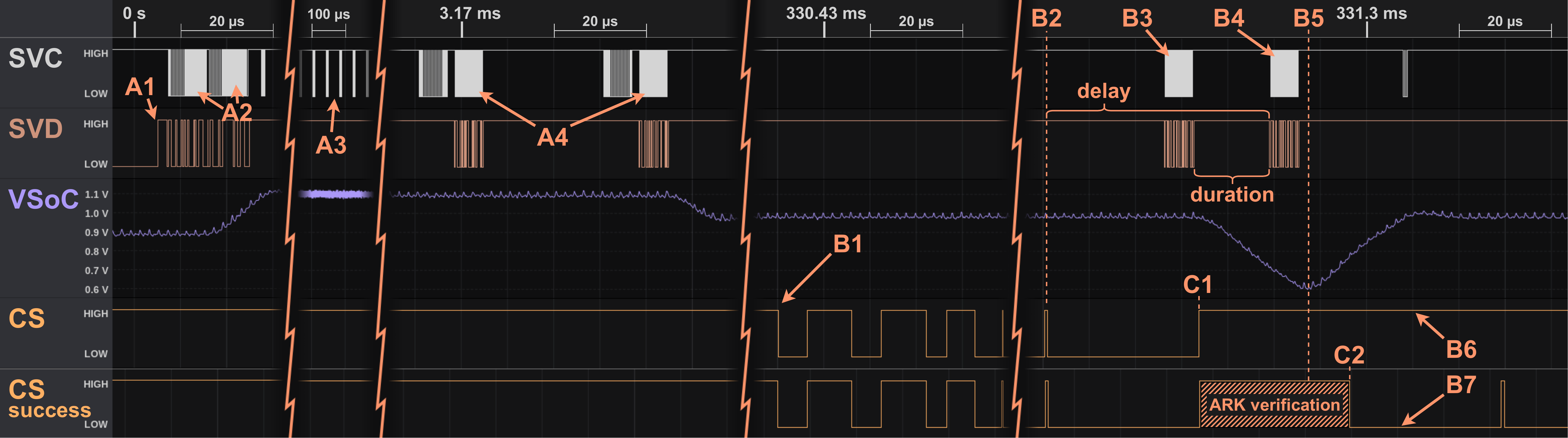}
    \caption{
Logic traces of a complete Attack Cycle including \ac{cs} traces of a successful and a failed attempt.
The ``A''-labels mark the \ac{svi2} bus activation, periodic telemetry reports, and disabling the telemetry reports, as described in Section \ref{sec:exp_setup:injection}.
Labels starting with ``B'' mark trigger events, the voltage drop injections, and the feedback mechanism described in Section \ref{sec:exp_setup:voltage-drop}.
The \ac{cs} edges marked with ``C'' are used to determine the initial window for the \emph{delay} parameter.
}
    \label{fig:complete-traces}
\end{figure*}

Figure~\ref{fig:complete-traces} depicts the output of the relevant signals of a successful glitch cycle.
In each cycle, the following steps are executed:
\begin{itemize}[leftmargin=1.2cm]
\item[A1] The Teensy detects the \ac{svi2} bus becoming active, starting the attack logic (\ref{sec:exp_setup:boot_detection})
\item[A2 - A4] To avoid later \ac{svi2} packet collisions, we inject two commands to disable the telemetry reports and set default voltages (\ref{sec:exp_setup:collisions}).
\item[B1 - B2] Using the number of \ac{cs} pulses and the \emph{delay} parameter we determined for the targeted CPU, we precisely trigger the voltage drop (\ref{sec:exp_setup:trigger}).
\item[B3 - B5] By injecting two \ac{svi2} commands (B3 and B4), we cause the voltage drop. The lowest voltage (B5) is determined by the \emph{duration} parameter (\ref{sec:exp_setup:voltage-drop}).
\item[B6 - B7] We observe further \ac{spi} traffic to distinguish between successful and failed attack attempts (\ref{sec:exp_setup:feedback}).
\end{itemize}
After each failed attempt, we start the next one by resetting the AMD \ac{soc} using the ATX reset line (see Figure \ref{fig:hardware-setup}).
Our attack cycle takes \num{3.14} ($\pm{}$ \SI{2}{\ms}) seconds, which amounts to just above 1100 attempts per hour.
This attack rate is limited by the ATX reset line timeout, which allows us to reset the AMD \ac{soc} only after around 3 seconds have passed since the last reset.

\subsection{SVI2 Bus Injection} \label{sec:exp_setup:injection}

On all AMD CPUs that we tested, the \ac{amd-sp} is powered by the \ac{soc} voltage rail, which is controlled by a dedicated \ac{vr} and a dedicated \ac{svi2} bus on CPUs with an SP3 socket~\cite{sp3_pinout}.
To inject packets onto this \ac{svi2} bus, we soldered two wires to its SVC and SVD lines.
While the bus is idle, both lines are permanently pulled to a logical high level by the CPU, which we use to inject packets by pulling the lines low.
We used an 8-channel open-drain driver (the LVC07A \cite{ti_lvc07a}) for this task.
Per bus line, we connected two channels of the driver in parallel to reliably achieve a logical low level accepted by the \emph{IR35204} \ac{vr}~\cite{internatio_ir35204_2016} present on our motherboard.

The driver's inputs are connected to one of the Teensy's I\textsuperscript{2}C hardware interfaces and are pulled high with a \SI{150}{\ohm} resistor.
Together with the Teensy's own open-drain drivers, this enables us to inject \ac{svi2} commands at a baudrate of \SI{4.6}{\Mbps}.
This is within the \SIrange{0.1}{21}{\Mbps} range commonly supported by the \acp{vr}~\cite{internatio_ir35201_2015, internatio_ir35204_2016, renesasel_isl62776_2020, richtekte_dualoutput_2019}, but faster than the \SI{3.3}{\Mbps} that we measured for our CPUs.

\subsubsection{\ac{svi2} Protocol} \label{sec:exp_setup:protocol}
The \ac{svi2} bus packet format is best described in \cite{renesasel_isl62776_2020} and \cite{richtekte_dualoutput_2019}. 
An \ac{svi2} command contains many configuration values, of which the following are of interest to us: The voltage domain selection bits, the \ac{vid} byte, the power state bits, and the \acf{tfn} bit.
All other values have a ``no change'' setting, which we choose for every injected packet.
Each \ac{svi2}-compliant \ac{vr} can regulate two voltage rails.
On motherboards with a single \ac{vr} (e.g., with AM4 socket~\cite{am4_pinout}), both the Core and \ac{soc} voltage rails (aka VDD and VDDNB, respectively) are regulated by that \ac{vr}.
The voltage domain selector bits are used to select which voltage rail is affected by an \ac{svi2} packet.
For Epyc CPUs, there is one \acl{vr} for each voltage rail.
Our experiments have shown that the Core (VDD) settings are used for both rails.

The \ac{vid} byte sets the main parameter of the \ac{vr}: the voltage of the selected voltage rail.
As there is no ``no change'' \ac{vid}, we must set a reasonable value every time we inject a command.
The default values we use for the Core and \ac{soc} voltage rails are the first values we observed on the bus.
The \acp{vr} use different power states for increased efficiency in low-power phases \cite{internatio_ir35201_2015, internatio_ir35204_2016, renesasel_isl62776_2020, richtekte_dualoutput_2019}.
We always choose the highest power state for our injections, as we noticed more significant voltage switching ripples in the lower power states, which cause our voltage faults to be less predictable.

\subsubsection{Boot Detection} \label{sec:exp_setup:boot_detection}
When the CPU starts its boot sequence (after a power on or a reset), there is a period when the \ac{vr} is already providing power to the CPU, but is not controlled via the \ac{svi2} bus \cite{internatio_ir35201_2015, internatio_ir35204_2016, renesasel_isl62776_2020, richtekte_dualoutput_2019}.
This period ends when the CPU signals the \ac{vr} to use the \ac{svi2} bus.
For all CPUs that we tested, the SVD line are constantly pulled low when the \ac{svi2} bus is inactive.
However, when the \ac{svi2} communication is activated, SVD transitions to a high state (A1 in Figure 
~\ref{fig:complete-traces}).
When the \ac{svi2} bus becomes deactivated again (e.g., when the \ac{soc} is reset), the SVD line constantly remains at a low level, which we use to arm our \ac{svi2} startup detection again.

\subsubsection{Avoiding Packet Collisions} \label{sec:exp_setup:collisions}
Once the \ac{svi2} bus is active, the CPU immediately sends two \ac{svi2} commands, configuring defaults for the two voltage rails (A2 in Figure \ref{fig:complete-traces}).
No more commands are sent on the \ac{svi2} bus until the \ac{ark} has been verified.
Therefore, we are not affected by interfering \ac{svi2} commands from the CPU during the packet injection.

In contrast, the periodic telemetry reports sent from the \ac{vr} to the CPU use the SVC line as a shared clock (A3 in Figure \ref{fig:complete-traces}).
This can cause packet collisions if left unattended.
To avoid possible interference with our packet injection, we disable the telemetry reporting shortly after the \ac{svi2} bus becomes active (A4 in Figure \ref{fig:complete-traces}).

\subsection{Voltage Drop} \label{sec:exp_setup:voltage-drop}

To lower the voltage level of the \ac{amd-sp}, we inject two commands into the \ac{svi2} bus (B3 and B4 in Figure \ref{fig:complete-traces}).
First, we configure a low \acf{vid} setting, and secondly, we inject the same \ac{vid} that was configured before the voltage drop (see \ref{sec:exp_setup:collisions}).
The voltage set by the first packet is too low for the \ac{amd-sp} to operate correctly and would cause non-recoverable errors, even if configured for only a short time.
However, due to the limited voltage regulation speed of the \ac{vr}, we inject the second command before the configured voltage is reached. 
This way, we can control the depth and shape of our voltage drop with only one parameter, the \emph{duration}.
Another advantage is that the voltage rail reaches its minimum for only a short moment, which we call the \emph{fault time} (B5 in Figure \ref{fig:complete-traces}).
The \emph{fault time} occurs directly after the second command injection, which allows us to trigger the fault injection precisely.

\subsubsection{Trigger} \label{sec:exp_setup:trigger}
As discussed in Section \ref{sec:attack_scenario:targeting_amd_sp}, counting the number of active low (negative) \ac{cs} pulses allows us to determine the time window for the \ac{ark} verification.
To more precisely control the \emph{fault time} (B5 in Figure \ref{fig:complete-traces}) within the \ac{ark} verification window, we use a \emph{delay} parameter, which is the time between the last counted \ac{cs} pulse and the \emph{fault time}.
Both timings are implemented on the Teensy using a busy loop where one iteration corresponds to \SI{12.5}{\ns}.
The complete trigger process proceeds as follows, see Figure~\ref{fig:complete-traces}:
\begin{itemize}
\item[A1] Starting with the boot detection, we count the number of \ac{cs} pulses.
\item[B1] The first \ac{cs} pulse is counted.
\item[B2] After counting the last \ac{cs} pulse\footnote{
To achieve a certain voltage drop depth within the \ac{ark} verification window, the first \ac{svi2} command has to be issued before the last \ac{cs} pulse (B3 Figure \ref{fig:complete-traces}).
In these cases we have to decrease the number of \ac{cs} pulses that we count.
}, we start the busy loop counter.
\item[B3] After (\emph{delay}\,${}-{}$\,\emph{duration}) busy loop cycles we inject the first \ac{svi2} command.
\item[B4] \emph{Duration} many busy loop cycles later -- exactly \emph{delay} busy loop cycles after B2 -- we inject the second \ac{svi2} command.
\item[B5] The \emph{fault time} is precisely determined by the \ac{cs} pulse count and \emph{delay}.
\end{itemize}

\subsubsection{Fault Feedback} \label{sec:exp_setup:feedback}

We can use the \ac{cs} line to infer what effect our voltage drop had on the execution of the \ac{amd-sp}.
Two different behaviors can be observed, see Figure~\ref{fig:ark_spi}:
\begin{itemize}
\item[B6] No further accesses to the SPI flash occur.
\item[B7] The \ac{amd-sp} continues to load data from the \ac{spi} flash.
\end{itemize}
For our attack firmware image with an invalid \ac{ark}, B6 means that the attack failed.
The reason is either that our key was correctly identified as invalid, or that we caused an unrecoverable fault in the \ac{rom} bootloader's operation.
In this case, the Teensy resets the target using the ATX reset line.
Since the \ac{rom} bootloader only continues to load data from the \ac{spi} flash when the loaded key was accepted as valid, B7 means that our attack succeeded.%

\subsection{Determining the Attack Parameters} \label{sec:exp_setup:parameters}

To successfully mount the glitching attack, we first need to determine the glitching parameters:
The \emph{delay}, responsible for the precise timing of our voltage drop, and the \emph{duration}, which sets the depth of the voltage drop (see Figure \ref{fig:complete-traces}).
As a first step, we limit both parameters to windows containing all sensible values (Sections \ref{sec:exp_setup:delay_window} and \ref{sec:exp_setup:duration_window}).
This is done manually using the serial interface of the Teensy, which took us around 30 minutes for each CPU.
These windows are then searched and refined using automated attacks (see Section \ref{sec:exp_setup:refining}).

\subsubsection{Delay Window} \label{sec:exp_setup:delay_window}
In the beginning, we limit the \emph{delay} parameter such that the \emph{fault time} always lies in the \ac{ark} verification window.
This is done by measuring the \ac{cs} line at \emph{fault time} for varying \emph{delay} parameters and firmware images.
With the \emph{duration} set to zero and an invalid \ac{ark} on the flash image, we can use the last \ac{cs} pulse to determine the first \emph{delay} value in the \ac{ark} verification window (C1 in Figure \ref{fig:complete-traces}).

Then we flash the original firmware image to the \ac{spi} flash.
Since this image's \ac{ark} is valid, our attack attempts -- with \emph{duration} set to zero -- will now always ``succeed''.
By again measuring the \ac{cs} line at \emph{fault time}, we now find the last \emph{delay} inside the \ac{ark} verification window (C2 in Figure \ref{fig:complete-traces}).
According to our observation, the resulting \emph{delay} window is about 2000 parameters wide. \todo{``parameters'' better than ``busy loop cycles''?}

\subsubsection{Duration Window} \label{sec:exp_setup:duration_window}
\begin{figure}[h]
    \includegraphics[width=\columnwidth, height=.5\columnwidth]{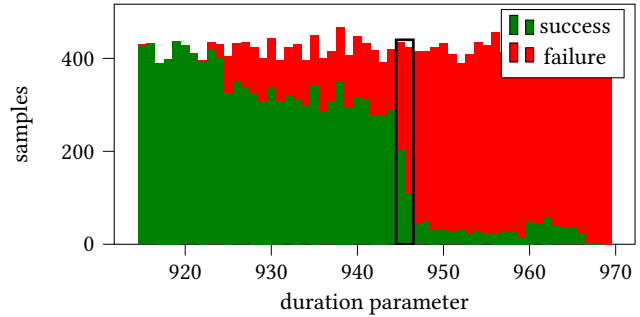}
    \vspace{-7mm}
    \caption{
Attack samples for \emph{duration} parameters in the transition window between always succeeding and always failing.
The attacks target the original firmware image on the AMD Epyc 72F3 CPU.
The final \emph{duration} window is marked in black and contains the values deemed most likely to cause a fault by the refinement process.
}
    \label{fig:duration-histogram}
\end{figure}

As a next step, we want to limit the \emph{duration} parameter, i.e., the voltage depth, so that we can search the resulting parameter space.
To do this, we use the already flashed original firmware image and run our attack with varying \emph{duration} parameters and a \emph{delay} that is inside the window specified above.
For shorter \emph{durations}, our attacks will mostly ``succeed'', but for longer \emph{durations}, it will transition to mostly ``failing'' (see Figure \ref{fig:duration-histogram}).
Using a binary search, we can identify the window of transition between these two extremes.

We expect to cause functional faults with \emph{duration} parameters inside this transition window, which our experiments confirm, see Figure~\ref{fig:duration-histogram}.
This observation aligns with other works that analyze voltage faults on ARM processors with respect to the depth and length of a voltage drop~\cite{timmers_controlling_2016}.

\subsubsection{Refining Parameters} \label{sec:exp_setup:refining}
To limit both parameters further, we repeatedly attempt our attack with randomly selected values from the two windows.
On each CPU we tested, it took us less than 6 hours to archive a first successful attempt.
The parameter space can now be limited further, e.g., to a window of $\pm 50$ \emph{delay} parameters and $\pm 10$ \emph{duration} parameters around the successful attempt's values.
With these smaller windows, we have an increased chance of achieving successes, which we use to limit the parameter space further.

\subsubsection{Results} \label{sec:exp_setup:results}
\begin{table}[h]
\begin{tabular}{r|cccc}
\textbf{Total}
& 72F3 (Zen 3) & 7272 (Zen 2) & 7281 (Zen 1) \\ \hline
Succ./Attempts &
170/486695 &
17/15459 &
144/110382 \\
Success Rate &
\SI{0.035}{\%} &
\SI{0.11}{\%} &
\SI{0.130}{\%}  \vspace{1em}\\
\textbf{Final Window} &&&\\
\hline
Succ./Attempts &
6/4653 &
6/3467 &
36/18309 \\
Success Rate &
\SI{0.129}{\%} &
\SI{0.173}{\%} &
\SI{0.197}{\%} \\
$\Delta$Delay/$\Delta$Dur. & 4/2 & 14/3 & 20/10 \\
\end{tabular}
\vspace{1mm}
  \caption{Attack results per CPU}\label{tbl:glitch-success}
\vspace{-1em}
\end{table}
We summarize the overall results in Table \ref{tbl:glitch-success}, together with the final parameter windows we used.
Our attack gains code execution reliably with an average waiting time between \SI{13.5}{\min} (Zen 1) and \SI{46.5}{\min} (Zen 3) for our final parameters.
However, the calculated success rates cannot be translated into a reliable worst-case time-to-exploit metric since the successful attempts are not uniformly distributed over time.

\subsection{Payloads\label{sec:attack}}
In this section, we present the attack payloads we executed leveraging the glitching of the \ac{amd-sp}'s \ac{rom} bootloader.
We briefly describe our approach to re-enable the attacks presented in~\cite{ccs19}.
For further details regarding these attacks, we refer to the original paper.

\paragraph{Dumping the \ac{rom} bootloader and extracting secrets}%
To analyze the endorsement key derivation process, we build a payload that extracts the \ac{rom} bootloader and SRAM contents of all targeted CPUs.
The payload writes the respective components to the \ac{spi} bus, including the \ac{vcek} secrets.
The \ac{cek} secrets were extracted from the \acf{ccp} using a similar payload.\footnote{
There is no public documentation available for the \ac{ccp}.
However, its functionality is described in the corresponding Linux kernel driver:~\cite{CCPDEV}.
}
In Section \ref{sec:key_derivation}, we use these secrets to derive the \ac{cek} and \ac{vcek} key of the exploited CPUs.

\paragraph{SEV Policy Override}%
In~\cite{ccs19}, the authors present attacks against \ac{sev}-protected \acp{vm} based on firmware issues present in the first generation of AMD Epyc CPUs (Zen 1). 
We successfully mounted these attacks on an AMD Epyc Zen 2 system, running the latest \ac{sev} firmware available from~\cite{AMD_SEV_FW}.
Similarly to~\cite{ccs19}, we patched the \ac{sev} firmware to ignore the guest's policy for the \texttt{DBG\_DECRYPT} command.
The target host was booted with a modified PSP OS firmware, which allowed us to update any \ac{sev} firmware signed with our own key.

\paragraph{AMD-SP Firmware Decryption}%
Our analysis of the \ac{amd-sp}'s firmware images for AMD Epyc Zen 3 CPUs showed that firmware components, such as the PSP OS and the SEV firmware, are encrypted, see Section~\ref{sec:background}.
In contrast to that, the analyzed AMD Epyc Zen 2 and Zen 1 images did not contain encrypted firmware images.
For AMD Epyc Zen 3 CPUs, we inferred the encryption mechanism by analyzing the \ac{rom} bootloader extracted from an AMD Epyc Zen 2 CPU.
Despite the fact that the Zen 2 firmware components were not encrypted, the \ac{rom} bootloader supports encrypted firmware files according to our anlysis.
To enable the firmware analysis on AMD Epyc Zen 3 systems and to better understand \ac{sev-snp}'s endorsement key derivation, we built a payload that extracts the firmware encryption key.
The firmware encryption scheme used in AMD CPUs is described in detail in the following section.

\section{Firmware Decryption\label{sec:firmware_decryption}}
The AMD-SP on Epyc Zen 3 CPUs uses AES in \acf{cbc} mode to decrypt firmware components stored on the external \ac{spi} flash.
Each component is prepended with a 256-byte header in the \ac{spi} flash.
The header contains meta-information about the respective component, such as a component's size and whether it is encrypted or not.
In case a component is encrypted, the header also contains the component's encryption key, denoted as \ac{ck} in the following text, and the \acf{iv} required for the decryption using $AES\text{-}CBC$. 
To protect the \ac{ck}, it is encrypted using AES in \acf{ecb} mode with a key stored within the \ac{amd-sp}'s filesystem on the \ac{spi} flash.
This key is referred to as \acf{ikek}~\cite{corebootPsp}.

Our analysis of the \ac{rom} bootloader of AMD Epyc Zen 2 CPUs revealed that the \ac{ikek} is encrypted, and the corresponding key, denoted as \acf{rk} in the following text, is held in non-readable memory areas of the \ac{ccp}.
There is no public documentation available for the \ac{ccp}.
However, its functionality is described in the corresponding Linux kernel driver, see~\cite{CCPDEV}.

In case a firmware component is encrypted, the \ac{amd-sp} on AMD Epyc CPUs performs the following steps:

\begin{enumerate}
  \item Load the \ac{ikek} from the \ac{spi} flash
  \item Decrypt the \ac{ikek} using the \ac{rk}:\\ $\rightarrow iKEK' = AES\text{-}ECB(rK,iKEK)$
  \item Decrypt the \ac{ck} using the decrypted \ac{ikek}:\\ $\rightarrow cK' = AES\text{-}ECB(iKEK',cK)$
  \item Decrypt the component using the decrypted \ac{ck}:\\ $\rightarrow plaintext = AES\text{-}CBC(cK', IV, data)$
\end{enumerate}

Using the glitch attack, we verified that the \ac{rk} is not directly accessible.
To analyze the firmware components on AMD Epyc Zen 3 CPUs, we created a payload that performs step 2, i.e., the \ac{ikek} decryption. 
With the decrypted \ac{ikek} ($iKEK'$), we could decrypt the PSP OS and the SEV firmware to enable further analysis of the \ac{vcek} key derivation process, which is presented in the following section.

\section{CEK \& VCEK Derivation} \label{sec:key_derivation}

Through the attacks presented in Section~\ref{sec:attack}, we have access to the firmware components that implement the key derivation for \ac{sev}'s endorsement keys and the corresponding secrets.
The \ac{cek} and \ac{vcek} are fundamental for the security properties of \ac{sev} (see Section~\ref{subsubsec:snp_remote_attestation}).
Both are derived from secret values burned into the fuses of the AMD \ac{soc}.
Each AMD \ac{soc} has a unique \ac{id} that can be used to retrieve certificates for the \ac{cek} and \ac{vcek} keys from AMD~\cite{AMD_CEK_KDS, AMDSevAPI}.

\subsection{Key Derivation Algorithms\label{sec:key_derivation:algos}}
In this section we present our analysis of the derivation algorithms for the \ac{cek} and the \ac{vcek}.

\subsubsection{\ac{cek} Derivation} \label{sec:key_derivation:cek}
The \ac{cek} is generated from a 32-byte secret.
This secret is expanded to 56 pseudorandom bytes using NIST's \emph{Key Derivation Function in Counter Mode} (KDF), specified in~\cite{chen_recommendation_2009}, with HMAC-SHA256 as \emph{Pseudorandom Function}.
The KDFs inputs for the \ac{cek} derivation are an empty context, the label ``sev-chip-endorsement-key'' and, as key, the SHA256 digest of the secret.
These 56 pseudorandom bytes can then be converted into an ECDSA key on the \emph{secp384r1} curve~\cite{brown_sec_2010}.
The algorithm used for this is NIST's \emph{Key Pair Generation Using Extra Random Bits}, specified in~\cite{technology_digital_2013}.

\subsubsection{\ac{id} Derivation} \label{sev:key_derivation:id}
The \ac{id} of the AMD \ac{soc} is generated from the same secret as the \ac{cek}.
This secret is interpreted as the private part of an ECDSA key on the elliptic curve \emph{secp256k1}, specified in~\cite{brown_sec_2010}.
The public part of this key, encoded as the concatenation of its two 32-byte coordinates, is then used as the \ac{id}.

\subsubsection{\ac{vcek} Derivation} \label{sec:key_derivation:vcek}
To derive the \ac{vcek}, a 48-byte secret value is used.
This secret is modified to incorporate the \acf{tcb} version string (see Section~\ref{subsubsec:snp_vcek}).
The \ac{tcb} version string consists of eight different one-byte \acfp{svn}, four of which are currently reserved (see Table~\ref{tab:tcb_version}).
We use $v_0, v_1, \dots, v_7$ to denote these \acp{svn} and $sec\_255$ to denote the initial secret.
To incorporate $v_0$ into this secret, we use $255 - v_0$ successive SHA384 operations:
$$
sec\_255
\overset{\text{\Tiny SHA384}}
\rightarrow
sec\_254
\overset{\text{\Tiny SHA384}}
\rightarrow
\dots
\overset{\text{\Tiny SHA384}}
\rightarrow
sec\_(v_0+1)
\overset{\text{\Tiny SHA384}}
\rightarrow
sec\_v_0
$$
By prefixing $sec\_v_0$ with eight zero-bytes and applying SHA384 again, the algorithm ``locks'' the first \ac{svn} and prepares it for the incorporation of the next \ac{svn}:
$$
sec\_v_0\_255 := SHA384(
\text{`\textbackslash0\textbackslash0\textbackslash0\textbackslash0\textbackslash0\textbackslash0\textbackslash0\textbackslash0'}
 \,||\; sec\_v_0)
$$
As suggested by this notation, we now apply $255 - v_1$ successive SHA384 operations to $sec\_v_0\_255$ to generate $sec\_v_0\_v_1$ and ``lock'' the second \ac{svn} using the prefix and SHA384 operation:
\begin{align*}
sec\_v_0\_255
\overset{\text{\Tiny SHA384}}
\rightarrow
sec\_v_0\_254
\overset{\text{\Tiny SHA384}}
\rightarrow
\dots
\overset{\text{\Tiny SHA384}}
\rightarrow
sec\_v_0\_v_1
\\
sec\_v_0\_v_1\_255 := SHA384(
\text{`\textbackslash0\textbackslash0\textbackslash0\textbackslash0\textbackslash0\textbackslash0\textbackslash0\textbackslash0'}
 \,||\; sec\_v_0\_v_1)
\end{align*}
This process is continued with the remaining six \acp{svn}, and we are left with the secret
$
sec\_v_0\_v_1\_v_2\_v_3\_v_4\_v_5\_v_6\_v_7
$, which is hashed one more time with SHA384 to obtain the final secret
\begin{equation}
\begin{aligned}
sec\_final &= SHA384(sec\_v_0\_v_1\_v_2\_v_3\_v_4\_v_5\_v_6\_v_7) \\
           (&= sec\_v_0\_v_1\_v_2\_v_3\_v_4\_v_5\_v_6\_(v_7-1)\,)\;.
\end{aligned}
\label{eqn:sec_final}
\end{equation}

We then use $sec\_final$ to generate the \ac{vcek} similarly to how the \ac{cek} is derived from its secret.
Using the label ``sev-versioned-chip-endorsement-key'' and the key $sec\_final$ as inputs to the same KDF as described in Section~\ref{sec:key_derivation:cek}, we again derive 56 pseudorandom bytes, which are turned into an ECDSA key on the \emph{secp384r1} curve using the same algorithm used in Section~\ref{sec:key_derivation:cek}.

\subsection{\ac{vcek} Design} \label{sec:key_derivation:design}

The goal of the \ac{vcek}, as described in Section~\ref{subsubsec:snp_vcek}, requires that from a given secret (e.g. $sec\_(v_0-1)$), we are not able to derive a secret for a higher \ac{svn} (e.g. $sec\_v_0$).
The cryptographic properties of SHA384 assure this, since SHA384 is practically infeasible to invert.

The \ac{sev-snp} API allows \ac{tcb} downgrades (see Section~\ref{subsubsec:snp_vcek}).
If, for example, we want to downgrade $sec\_final$'s last \ac{svn} by one, we can apply one SHA384 operation to $sec\_final$ and generate the ECDSA key from the resulting secret.
However, this mechanism can only be used to downgrade the last \ac{svn}.
To allow downgrades of all \acp{svn}, the \ac{sev} firmware has access to all of the secrets in~(\ref{eqn:all_secrets}).
\begin{equation}
\begin{aligned}
&sec\_(v_0-1)\\
&sec\_v_0\_(v_1-1)\\
&sec\_v_0\_v_1\_(v_2-1)\\
&\dots\\
&sec\_v_0\_v_1\_v_2\_v_3\_v_4\_v_5\_v_6\_(v_7-1) = sec\_final
\end{aligned}
\label{eqn:all_secrets}
\end{equation}
For example, to derive the \ac{vcek} for the \ac{tcb} version string
$$(v_0, v_1 - 2, v'_2, \dots, v'_7)\;,$$
we can apply one SHA384 operation to $sec\_v_0\_(v_1-1)$ and then continue the \ac{vcek} derivation algorithm with the \acp{svn} $v'_2, \dots, v'_7$.
A potential issue is that we can choose values for $v'_2, \dots, v'_7$, which are higher than the original values $v_2, \dots, v_7$.
We can, for example, derive the secret
\begin{equation}
sec\_v_0\_(v_1-2)\_0\_0\_0\_0\_255\_254 \; , \label{eqn:down_upgrade}
\end{equation}
which would result in a valid \ac{vcek} with the \ac{svn} 255 for both the \ac{sev} application and \si{\micro}Code patch level.
However, this does not constitute a security vulnerability as the downgraded \ac{svn} belongs to an insecure firmware component with a higher privilege level than the firmware components with upgraded \acp{svn}.
In the example above, the \ac{svn} $v_1 - 2$ refers to an insecure \ac{psp-os} firmware, whose security vulnerabilities could potentially be used to leak the secret~(\ref{eqn:down_upgrade}).

\subsection{Implementation on the \ac{amd-sp}} \label{sec:key_derivation:implementation}

Both the \ac{id} and the \ac{cek} derivation algorithms described above are implemented by the \ac{sev} application.
Their shared secret value is derived by the \ac{rom} bootloader, which passes the secret to the \ac{psp-os} in a readable buffer of the \ac{ccp}.
The \ac{sev} application can then access the secret through the syscall interface of the \ac{psp-os}.

The \ac{sev} application is also responsible for deriving the \ac{vcek}, but the secrets derivation algorithm is split up between \ac{rom} bootloader, the \ac{psp-os}, and the \ac{sev} application.
The \ac{rom} bootloader derives the initial \ac{vcek} secret $sec\_255$ from the fuses of the \ac{amd-sp}.
The first \ac{svn}, i.e., the \ac{svn} labeled \texttt{BOOT\_LOADER} in Table~\ref{tab:tcb_version}, is part of the header of the \ac{psp-os} binary on the SPI flash.
The \ac{rom} bootloader uses this first \ac{svn} $v_0$ to derive the secrets:
\begin{equation}
sec\_v_0\_255 \quad\text{and}\quad sec\_(v_0 - 1)\;.
\label{eqn:bl-seeds}
\end{equation}
Once it has verified the \ac{psp-os} signature (the header is included in this signature), both secrets~(\ref{eqn:bl-seeds}) are passed onto the \ac{psp-os}.

According to the \ac{sev-snp} API specifications~\cite{AMDSNPAPI}, the second \ac{svn} corresponds to the ``trusted execution environment.''
In the firmware image we analyzed, \todo{Mention which firmware image we used.} the \ac{psp-os} binary was responsible for the boot process and acted as an operating system running on the \ac{amd-sp}.
As a result, this second \ac{svn} and the next four ``reserved'' \acp{svn} are set to zero by the \ac{psp-os}.
Every time a \ac{sev} application is loaded, its \ac{svn}, $v_6$, and the hardcoded \acp{svn}, $v_1$ to $v_5$, are used to derive all the first seven secrets of~(\ref{eqn:all_secrets}) and
$$
sec\_v_0\_\dots\_v_6\_255 \;.
$$
These secrets are then passed to the \ac{sev} application, which incorporates the last \ac{svn} -- the \si{\micro}Code patch level.

\subsection{CEK/VCEK Derivation and Fault Attacks}
With the glitching attack presented in Section~\ref{sec:exp_setup}, an attacker can not only execute an arbitrary \ac{psp-os} firmware component, but also, choose an arbitrary \ac{svn} for its header.
By creating a payload with the highest possible \ac{svn}, $255$, the attacker forces the \ac{rom} bootloader to derive the secrets
$$ sec\_255\_255 \quad\text{and}\quad sec\_254\;. $$
We extracted these secrets, together with the \ac{cek} secret, using payloads described in Section~\ref{sec:attack}.
With these secrets and the algorithms described in Section~\ref{sec:key_derivation:algos}, we were able to derive the \acp{cek} and \acp{id} for all our target CPUs, as well as the \acp{vcek} of our AMD Epyc Zen 3 CPU for all possible TCB component versions.

\section{Discussion\label{sec:discussion}}
In this section we evaluate the feasibility and impact of our attack, and propose potential mitigations.

\subsection{Attack Evaluation\label{subsec:attack_evaluation}}

To evaluate the real-world applicability of our attack, we compare an attacker's capabilities with the attack requirements.
We focus on attack scenarios relevant in cloud environments, as presented in Section~\ref{sec:attack_scenario}.

For the ``Debug Override'' attack, the attacker first replaces the original \ac{sev} firmware with a custom firmware and re-signs the firmware image, see Section~\ref{sec:attack}.
To mount the glitching attack described in the previous sections, the attacker then prepares the target host as described in Section~\ref{sec:exp_setup}.
As the Teensy µController is small enough to fit into a standard server enclosure, the physical setup introduces no additional requirements regarding the installation into the datacenter.
This initial preparation, including the determination of the glitching parameters, does not pose a serious challenge for the attacker. 
We were able to prepare our target system for attacking the Epyc 72F3 in under four hours.

The attacker will then boot the target system, and the µController will perform the actual glitching via the \ac{svi2} packet injection. 
The µController will automatically reset the target if the glitch attempt failed, leading to an increased boot time of the target.
Once the target is fully booted, the attacker can leverage \ac{sev}'s debug API to decrypt a \ac{vm}'s memory. 
As described in Section~\ref{sec:exp_setup}, a attack machine is required to control the µController.
While in a cloud scenario, a neighboring host could be used to control the µController.
However, with minor firmware modifications, the Teensy µController would be capable of automatically performing the glitching attack on its own.
As the manipulation of the \ac{sev} firmware image does not change a \ac{vm}'s measurement, the validation of attestation reports will succeed even though the target host does not execute a genuine AMD \ac{sev} firmware.

For the second scenario, instead of preparing a system to host the targeted \acp{vm}, the attacker prepares and then extracts the endorsement key of an arbitrary \ac{sev} capable CPU.
The endorsement key is then used to fake attestation reports. 
In the case of \ac{sev-snp}, the faked attestation report allows an attacker to migrate the victim \ac{vm} to a host with a malicious \ac{ma}, see Section~\ref{subsubsec:snp_migration}.
Using the exported \acf{oek}, the malicious \ac{ma} can decrypt the victim's memory pages.
For pre-\ac{sev-snp} targets, the extracted endorsement key allows to mount the migration attack as described in~\cite{ccs19}.
Compared to the first scenario, this approach relaxes the requirements for the attacker as the glitching attack can be performed in a controlled environment,	and the extracted keys can then be used to target remote systems.

Either attack scenario poses a threat for \ac{sev}-protected \acp{vm}, as they can be carried out by insider attackers such as system administrators and require only cheap and easily available hardware.

\subsection{Implications for the SEV Ecosystem\label{subsec:discussion_implications}}
The two attack scenarios presented in Section~\ref{sec:attack_scenario} allow an attacker to overcome SEV's protection guarantees. 
However, for the first scenario, the attacker must have physical access to the system running the targeted \ac{vm}.
While it is possible to migrate the target \ac{vm} to a host under the attacker's control, the requirement to have physical access to the targeted system still poses a challenge for the attacker given that modern data centers employ several physical security measures such as access control and 24h surveillance.

In contrast, the extraction of \ac{sev}'s endorsement keys allows an attacker to create valid attestation reports and requires only physical access to an arbitrary \ac{sev}-capable CPU.
The \ac{sev}-technology offers no mechanism to limit the lifetime of the endorsement keys.
Even with the \acf{vcek}, introduced with \ac{sev-snp}, the endorsement keys are still built on a chip-unique secret that, once extracted, can be used to derive all possible \acp{vcek} for that CPU.

The attestation reports play a central role in the trust-model of \ac{sev}.
They provide the \ac{vm} owner with the guarantee that the \ac{vm} was not tampered with during deployment and that the remote host uses a genuine AMD CPU with \ac{sev} protection in place.

By extracting the endorsement keys, we showed that a valid signature over \ac{sev} attestation reports is not sufficient to prove that the report originates from an authentic AMD system.
Without trusting the remote party, \ac{vm} owners cannot verify the integrity of their \ac{vm} or the associated \acf{ma}.

Thus, based on the results presented in this paper, the remote attestation feature of \ac{sev} must be considered broken on Zen 1, Zen 2, and Zen 3 AMD Epyc CPUs.

\subsection{Potential Mitigations}
We see two different strategies that can be pursued to mitigate our attack.
On the one hand, one could try to prevent the adversary from achieving code execution (Section ~\ref{sec:discussion:mitigations:hwsw}).
On the other hand, one could try to protect the architecture keys from being extracted, even if the adversary manages to achieve code execution (Section~\ref{sec:discussion:mitigations:arch}).

\subsubsection{Prevent adversarial code execution\label{sec:discussion:mitigations:hwsw}}

The threat of fault injection for gaining adversarial code execution can be tackled from different directions.
One could try to detect malicious voltage drops/glitches, and as a consequence, shut down the system to prevent further damage.
Alternatively, one could try to prevent faulty execution in the presence of glitches, for instance, by introducing redundancy.
Both approaches might imply the need for changes in the hardware or software design.

\paragraph{Hardware-based detection/prevention}

Voltage monitoring circuits -- as commonly implemented in modern smartcards -- could help to detect glitches.
A recent patent by NVIDIA \cite{rajpathak_cross_2020} proposes a cross-domain voltage glitch detection circuit, which can be implemented into a \ac{soc}.
The main idea is that circuits in different independent voltage domains monitor the voltage levels in the other domains, and if there is a glitch on a specific rail, an alert signal is asserted.
In our opinion, this is a promising approach.
However, it should be kept in mind that there might exist voltage glitch shapes that can cause faulty behavior but can not be detected by a particular protection circuit.

We share the opinion with the authors of the \emph{VoltPillager} attack~\cite{voltpillager} that voltage glitching can not be prevented by protecting the \ac{vr} bus/protocol through cryptographic authentication.
Furthermore, although the \ac{vr} bus is an easy access point for injecting voltage glitches, the adversary could also inject glitches by other means, e.g., by altering the PWM signal output of the \ac{vr} or entirely replacing the \ac{vr} with a custom injection setup.

One might think that fully integrating the \ac{vr} into the \ac{soc} could be the ultimate solution.
However, faults can not only be induced by glitching the supply voltage.
In the past couple of years, \ac{em} fault injection techniques against modern CPUs have been examined to inject faults in a more targeted and contactless way~\cite{trouchkine_fault_2020, trouchkine_electromagnetic_2021}.
Consequently, a holistic view is necessary to prevent all kinds of fault injection techniques that can lead to code execution on the target device.

\paragraph{Software-based detection/prevention}
Hardening the \ac{rom} bootloader might be another option to prevent the adversary from gaining code execution.
However, this is a complex task since the characteristics and potentials of faults are not well understood.
Particularly, there is no model which covers all possible faults.

Though, there are generic countermeasures that can decrease the probability of successful attacks~\cite{witteman_riscure_2018, ban_arm_2020}.
For instance, constants with large hamming distance make it hard to flip one valid value to another, double checks protect branch conditions, loop integrity checks make sure that the loop exits as intended, and a global counter can be used to monitor the program flow and detect anomalies.
For assessing software countermeasures against fault attacks, different simulation-based frameworks  have been proposed~\cite{holler_qemubased_2015, schirmeier_fail_2015}.
In our opinion, the general approach of software-based mitigations is promising because they can protect not only against fault attacks by voltage glitching.
Nevertheless, these countermeasures come at the cost of some overhead in execution time, and therefore, performance reduction.

\subsubsection{%
	Prevent key extraction%
	\label{sec:discussion:mitigations:arch}}

One of the key insights from~\cite{ccs19} is that the \ac{cek} should depend on TCB components versions.
In this case, \acp{cek} extracted using firmware issues are no longer valid after the firmware has been updated.
With \ac{sev-snp}, a similar model has been adopted by AMD in the form of the \ac{vcek}.

The \ac{vcek} is derived from a secrets that depend on TCB component versions.
However, as we have shown in this work, this dependency still allows an attacker to extract valid \acp{vcek} for all possible TCB versions.
Although the \ac{vcek} is bound to firmware versions, it does \emph{not} depend on the \emph{functionality} of the respective firmware component.
The dependency between the firmware version, represented by a field in a component's header, and the its functionality is only implicit.
Components, including their firmware version field in the header, are signed by the \ac{ark}.
A valid signature links the component to it's firmware version in the signed header.

To increase the security properties of the \ac{vcek} key derivation, we propose to include a hash -- instead of the firmware version -- of the respective TCB component in the secrets, similar to the DICE model proposed by the Trusted Computing Group~\cite{tcg_dice}.
Including the hash binds a secret to the component's functionality. 
Using a key-derivation function, the \ac{rom} bootloader binds the $sec\_255$ secret to the hash of the PSP OS, see Section~\ref{sec:key_derivation:implementation}.

In this model, if an attacker could get code execution on the \ac{amd-sp}, the secrets that could be extracted are useless as they depend on the attacker's payload.

While this model presents a security improvement compared to the current version of the derivation model, it also has practical drawbacks: any \emph{functional} update of a component, i.e., a non-security-related change, still results in a new \ac{vcek}. 
In addition to this, this model prevents TCB rollback of endorsement keys as currently supported by \ac{sev-snp}.
Without knowing the specific requirements of the \ac{amd-sp}'s firmware components, it is impossible to fully evaluate the applicability of the proposed model.
Nevertheless, from a pure security perspective, we believe including the hash within the intermediate secrets provides a strengthened security model.

\section{Conclusion\label{sec:conclusion}}

The attacks presented in this paper highlight \ac{sev}'s insufficient protection against physical attacks.
Despite its crucial role for \ac{sev}'s security properties, the \ac{amd-sp} can be tricked into executing attacker-controlled code.
The hardware setup to mount the presented glitching attack is cheap and readily available.
Building on this setup, we presented how an adversary with physical access to the target host can implant a custom \ac{sev} firmware that decrypts a \ac{vm}'s memory using \ac{sev}'s debug API calls.

Furthermore, we showed that the glitching attack enables the extraction of endorsement keys.
The endorsement keys play a central role in the remote attestation mechanism of \ac{sev} and can be used to mount software-only attacks.
Even an attacker without physical access to the target host can use extracted endorsement keys to attack \ac{sev}-protected \acp{vm}.
By faking attestation reports, an attacker can pose as a valid target for \ac{vm} migration to gain access to a \ac{vm}'s data.
The severity of the presented software-only attacks is amplified by the fact that an attacker can perform the key extraction on an AMD CPU unrelated to the CPU hosting the targeted VM, i.e., on an AMD Epyc CPU bought by the attacker for the sole purpose of extracting an endorsement key.

Our analysis revealed that the TCB versioning scheme introduced with \ac{sev-snp} does not protect against the presented attacks.
Based on our results, we conclude that \ac{sev} cannot adequately protect confidential data in cloud environments from insider attackers, such as rogue administrators.
The presented attacks do not rely on firmware issues and can not be easily mitigated.
Hence, we proposed mitigations for future AMD Epyc CPUs.
Nevertheless, to the best of our knowledge, all AMD Epyc CPUs of the Zen 1, Zen 2 and Zen 3 microarchitectures are susceptible to the presented attacks.

\bibliographystyle{ACM-Reference-Format}
\bibliography{paper}

\end{document}